\documentclass[10pt,a4paper]{article}
\usepackage{amsmath,amssymb}
\usepackage{graphicx}
\usepackage{subfigure}
\usepackage{fancyhdr}
\usepackage[textwidth=14cm,textheight=22cm]{geometry}
\begin{document}
\title{\Large{\textbf{A New Method of Analyzing Eigenvalues and Eigenfunctions of Rosenbluth Collision Operator}}}
\author{\large{Kaifeng Chen}}
\date{\large{10.2012}}
\maketitle
\section{\Large{Introduction}}
The problems in plasma physics always require some analysis of the collisions. An equation which has a general form describing collisions is the "Boltzmann Equation". It has the form as follows:
\begin{equation}
 \frac{\partial f_{\alpha}}{\partial t}+\textbf{v}\cdot\frac{\partial f_{\alpha}}{\partial \textbf{x}}+\textbf{a}\cdot\frac{\partial f_{\alpha}}{\partial \textbf{v}}=\large(\frac{\partial f_{\alpha}}{\partial t}\large)_c
\end{equation}
Where $f_{\alpha}(\textbf{x},\textbf{v},t)$ is the Boltzmann function of $\alpha$ particle in the usual notation, $\textbf{a}$ is the macroscopic contribution to the acceleration of the particle at $(\textbf{x},\textbf{v},t)$ and the term on the right represents collisions. A commonly used form of the collision term is of the Fokker-Planck type\cite{1}
\begin{equation}
 \large(\frac{\partial f_{\alpha}}{\partial t}\large)_c=\frac{\partial}{\partial v_i}\large\{-A_i f_{\alpha}+\frac{1}{2}\frac{\partial}{\partial v_j}(B_{ij}f_{\alpha})\large\}
\end{equation}
Here $A_i$ and $B_{ij}$ are the "friction" and "diffusion" coefficients respectively. Several different kinds of forms of $A_i$ and $B_{ij}$ are provided by previous plasma physicists. A widely used kind of collision term is called "Rosenbluth Collision Term"\cite{2}, which can be written in the following form:
\begin{equation}
 \large(\frac{\partial f_{\alpha}}{\partial t}\large)_c=-\sum_{\beta}\Gamma_{\alpha\beta}\frac{\partial}{\partial \textbf{v}}\cdot\big(f_{\alpha}\frac{\partial H_{\alpha\beta}}{\partial \textbf{v}}-\frac{1}{2}\frac{\partial}{\partial \textbf{v}}\cdot \frac{\partial^2 G_{\alpha\beta}}{\partial \textbf{v}\partial\textbf{v}}\big)
\end{equation}
In this collision term, $H_{\alpha\beta}$ and $G_{\alpha\beta}$ are called Rosenbluth potentials. They satisfy following relations:
\begin{subequations}
\begin{align}
 H_{\alpha\beta}(\textbf{v})&=\frac{m_{\alpha}+m_{\beta}}{m_{\beta}}\int d\textbf{v'}\frac{f_{\beta}(\textbf{v'})}{|\textbf{v-v'}|}\\
 G_{\alpha\beta}(\textbf{v})&=\int d\textbf{v'}f_{\beta}(\textbf{v'})|\textbf{v-v'}|\\
 \nabla^2_{\textbf{v}}H_{\alpha\beta}&=-\frac{4\pi(m_{\alpha}+m_{\beta})}{m_{\beta}}f_{\beta}\\
 \nabla^2_{\textbf{v}}G_{\alpha\beta}&=\frac{2m_{\beta}}{m_{\alpha}+m_{\beta}}H_{\alpha\beta}
\end{align}
\end{subequations}
Several other researchers have paid much attention about the eigenvalues and eigenfunctions of this collision term. Lenard and Berntein(LB)\cite{3} reduced the collision diffusion coefficient to be a constant related only to thermal velocity[Eq.\eqref{eq:1}]. Then they got an exact analytic solution with a dispersion relation. After their work, C.S.Ng,A.Bhattacharjee and F.Skiff\cite{4} further discussed the completeness of LB collision term theoretically. Their results indicated that the LB operator has a discrete eigenmode spectrum,totally different from Case-Van Kampen eigenmodes\cite{5}\cite{6}.
\begin{equation}\label{eq:1}
 \large(\frac{\partial f_{\alpha}}{\partial t}\large)_c=\nu\frac{\partial}{\partial v_i}\large(v_if_{\alpha}+v_0^2\frac{\partial f_{\alpha}}{\partial v_i}\large)
\end{equation}
J.P.Dougherty\cite{1} derived his own kinetic equation based on the conservation theorems[Eq.\eqref{eq:2}]. Then M.W.Anderson and T.M.O'Neil\cite{7} analytically gave a theoretical formulation of all the eigenvalues and eigenfunctions of the Dougherty collision term. \\
\begin{equation}\label{eq:2}
 \large(\frac{\partial f_{\alpha}}{\partial t}\large)_c=\nu\frac{\partial}{\partial v_i}\large\{(v_i-u_i)f_{\alpha}+\frac{KT}{m_{\alpha}}\frac{\partial f_{\alpha}}{\partial v_i}\large\}
\end{equation}
However, the previous researchers could only deal with Fokker-Planck collision term by reducing its complexity and difficulty to a much simpler mathematical form. For a general case, there are still no thorough and detailed discussions.\\
In this paper, we utilize Rosenbluth collision term for only ion-ion collisions and discuss the eigenvalues and eigenfunctions of this operator. Section 2 mainly focuses on the mathematical deducing procedure of linearizing Rosenbluth collision term. By comparing the whole operator and the differential term in the operator, Section 3 discusses the minimum eigenvalues of the least damped wave modes and their corresponding eigenvalues. Section 4 concentrates mainly on the theoretical analysis of the differential term including a proof of the completeness of the eigenvalues and orthogonality of the eigenfunctions calculated only by this term. Section 5 gives further discussions about the differential term. Some mathematical transformations are used and the continuum of the corresponding eigenvalues spectrum is indicated.Section 6 sums up the all conclusions.
\section{\Large{Linearization of Rosenbluth Collision Term}}
For research convenience and experiment reality, Fokker-Planck equation with this collision operator is always linearized to its first order perturbation by assuming the zeroth distribution function as the a Maxwellian form $f^M$\cite{8}.
\begin{subequations}
\begin{align}
 f&=f^M+f^1e^{i(kz-\omega t)}\\
 E&=-\Delta \phi,where~\phi=\phi_0+\phi_1e^{i(kz-\omega t)}
 \end{align}
\end{subequations}
After that procedure, one can get:
\begin{equation}
 \large(\frac{\partial f^11}{\partial t}\large)_c=i(kv_z-\omega)f^11+i\frac{Ze\phi_1kv_z}{T}f^M=C[f^M,f^1]+C[f^1,f^M]
\end{equation}
The linearized collision operator includes two parts: $C[f^M,f^1]$,integral part and $C[f^1,f^M]$, the differential part. These two parts can be written in the form:
\begin{subequations}
\begin{align}
 C[f^M,f^1]&=-\Gamma\frac{\partial}{\partial \textbf{v}}\cdot\big(f^M\frac{\partial H_{01}}{\partial \textbf{v}}-\frac{1}{2}\frac{\partial}{\partial \textbf{v}}\cdot f^M\frac{\partial^2 G_{01}}{\partial \textbf{v}\partial \textbf{v}}\big)\label{eq:3}\\
 C[f^1,f^M]&=-\Gamma\frac{\partial}{\partial \textbf{v}}\cdot\big(f^1\frac{\partial H_{10}}{\partial \textbf{v}}-\frac{1}{2}\frac{\partial}{\partial \textbf{v}}\cdot f^1\frac{\partial^2 G_{10}}{\partial \textbf{v}\partial \textbf{v}}\big)\label{eq:4}
\end{align}
\end{subequations}
The collision terms in Eq.\eqref{eq:3} and Eq.\eqref{eq:4}can be simplified by working in spherical coordinates:$v=|\textbf{v}|,\mu=v_z/v$,and by using symmetry properties of $f^M$ and $f^1$,
\begin{equation}\label{eq:5}
 \hat C_{01}f^1=C[f^M,f^1]=\Gamma[\frac{n_i}{(2\pi)^{3/2}v_i^3}]^2\large(4\pi f^Mf^1-\frac{H_{01}}{2}f^M+\frac{v^2}{2}\frac{\partial^2G_{01}}{\partial v^2}f^M\large)
\end{equation}
\begin{equation}\label{eq:6}
\begin{split}
 \hat C_{10}f^1=C[f^1,f^M]=&\Gamma[\frac{n_i}{(2\pi)^{3/2}v_i^3}]^2\big\{-\frac{1}{2v^2}\frac{\partial}{\partial v}[v\frac{\partial H_{10}}{\partial v}e^{-v^2/2}\frac{\partial}{\partial v}(e^{v^2/2}f^1)]\\
 &+\frac{1}{2v^3}\frac{\partial G_{10}}{\partial v}\frac{\partial}{\partial \mu}
 [(1-\mu^2)\frac{\partial f^1}{\partial \mu}]\big\}
\end{split}
\end{equation}
$(H_{01},G_{01})$ and $(H_{10},G_{10})$ refer to Rosenbluth potentials of perturbation $f^1$ and background $f^M$,respectively. The collision strength parameter $\Gamma$,$H_{10}$ and $G_{10}$ are given by
\begin{subequations}
\begin{align}
 \Gamma&=\frac{4\pi Z^4e^4}{m_i^2}ln\Lambda=\frac{3\sqrt{\pi}v_i^4}{\lambda n_0}\\
 \frac{\partial H_{10}}{\partial v}&=2(2\pi)^{3/2}\frac{\sqrt{2/\pi}ve^{-v^2/2}-erf(v/\sqrt{2})}{v^2}\label{eq:7}\\
 \frac{\partial G_{10}}{\partial v}&=\frac{1}{2}\frac{\partial H_{10}}{\partial v}+(2\pi)^{3/2}erf(v/\sqrt{2})\label{eq:0}
\end{align}
\end{subequations}
Where $erf(x)$ is the error function. To make analysis more convenient, we simply make the normalizations, $f^M\rightarrow f^M[n_i/(2\pi)^{3/2}v_i^3]^2$ and $f^1\rightarrow f^1[n_i/(2\pi)^{3/2}v_i^3]^2$and denote $\hat C_{10}$ as the differential part of the operator and $\hat C_{01}$ as the integral part.\\
in reference \cite{8}and\cite{9}, the perturbed ion distribution function $f^1$ is written as a sum of Legendre polynomials $P_l$,
\begin{equation}\label{eq:14}
 f^1=\sum_{l=0}^{+\infty}a_l(v)P_l(v)
\end{equation}
The expansion by using Legendre polynomials is also called "Moment Expansion".If there is no electric field and the perturbation is isotropic both in position space and velocity space, the equation of individual $a_l$ will be decoupled. This makes analysis of collision term much more easier, and that is why moment expansion is so widely used.
\section{\Large{Eigenvalues and Eigenfunctions}}
For further discussion, here we only consider $f^1$ to be isotropic in position place and velocity space and all the calculations are under the assumption of no electronic field. Then the Fokker-Planck equation becomes:
\begin{equation}\label{eq:7}
\begin{split}
 &\large\{\frac{-3\sqrt{\pi}}{v^2k\lambda}[\frac{\partial}{\partial v}(\frac{v}{2}\frac{\partial H_{10}}{\partial v}e^{-v^2/2}\frac{\partial}{\partial v}[e^{v^2/2}a_l])+\frac{1}{2v}\frac{\partial G_{10}}{\partial v}l(l+1)a_l]\large\}\\
 &+\big[\frac{3\sqrt{2}f^Ma_l}{k\lambda}-\frac{3f^M}{2^{3/2}\pi k\lambda}(\frac{H_l}{2}-\frac{v^2}{2}\frac{\partial^2 G_l}{\partial v^2})\big]=-i\omega a_l
\end{split}
\end{equation}
And $H_l$ and $G_l$ are Rosenbluth coefficients:
\begin{subequations}
\begin{align}
 H_l(v)&=\frac{8\pi}{(2l+1)v}\big(\frac{1}{v^l}\int_0^va_l(v')v'^{2+l}dv'+v^{l+1}\int_v^{+\infty}
 a_l(v')v'^{1-l}dv'\big)\\
\begin{split}
 \frac{\partial^2 G_l(v)}{\partial v^2}&=\frac{-4\pi}{4l^2-1}\int_0^va_l(v')v'^{2+l}\big(\frac{l(l-1)}{v^{l+1}}-\frac{(l+1)(l+2)(l-1/2)v'^2}{(l+3/2)v^{l+3}}\big)dv'\\
 &-\frac{4\pi}{4l^2-1}\int_v^{+\infty}\frac{a_l(v')}{v'^{l-3}}\big(l(l-1)v^{l-2}
 -\frac{(l+1)(l+2)(l-1/2)v^l}{(l+3/2)v'^2}\big)dv'
\end{split}
\end{align}
\end{subequations}
Eq.\eqref{eq:7}is the eigenvalue equation. The first term in \{~\} in Eq.\eqref{eq:7} is form the differential part and the second term in [~] comes from the integral part.\\
In order to calculate the eigenfunction $\lambda$ of Eq.\eqref{eq:7}, a common method by expanding $a_l$ in Sonine polynomials $S_n^{l+1}(v^2/2)$ has been discussed by previous researchers\cite{8}\cite{9}. But in our work, numerical simulation is employed by finite differncing the derivative terms and discretizing the integrals on velocity grid $v_i = \{v_1,v_2,\cdots,v_{j max}\}$, where $v_{jmax}$ is the maximum velocity used in simulation and it is normalized to thermal velocity. $j max$ is the maximum number of velocity girds.
By comparison, the eigenvalues of the differential operator are also calculated. As to our results, only the minimum eigenvalues are taken into consideration because their corresponding eigenmodes are of least resistance. And it is very obvious from Eq.\eqref{eq:7} that all eigenvalues $\omega$ should be imaginary numbers. Therefore, in our discussion, only the imaginary parts of eigenvalues are taken into account.
The results are shown as follows:
\begin{figure}[!h]
\centering
\subfigure[Figure 1]{
\includegraphics[width=6.5cm,height=4.5cm]{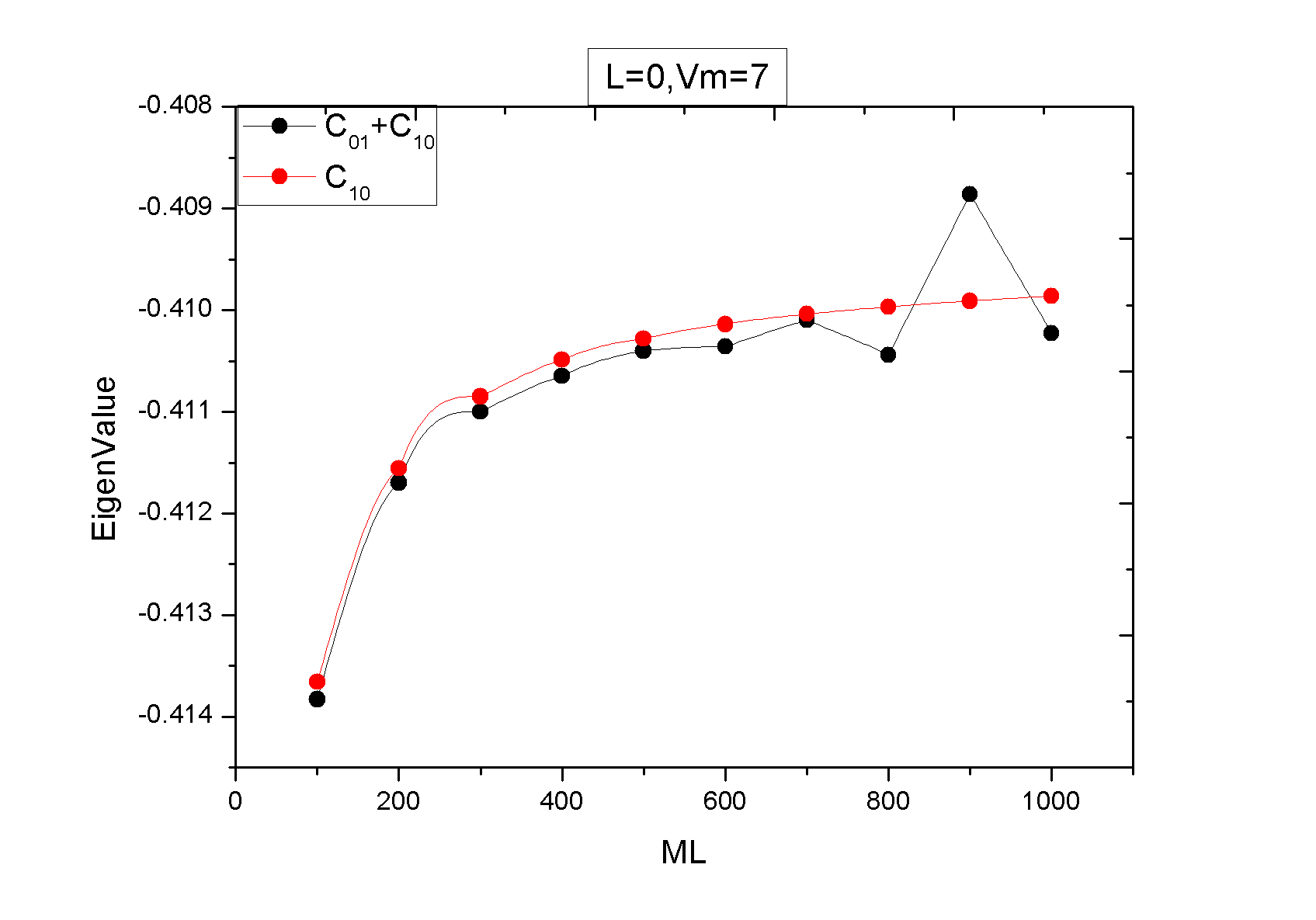}}
\subfigure[Figure 2]{
\includegraphics[width=6.5cm,height=4.5cm]{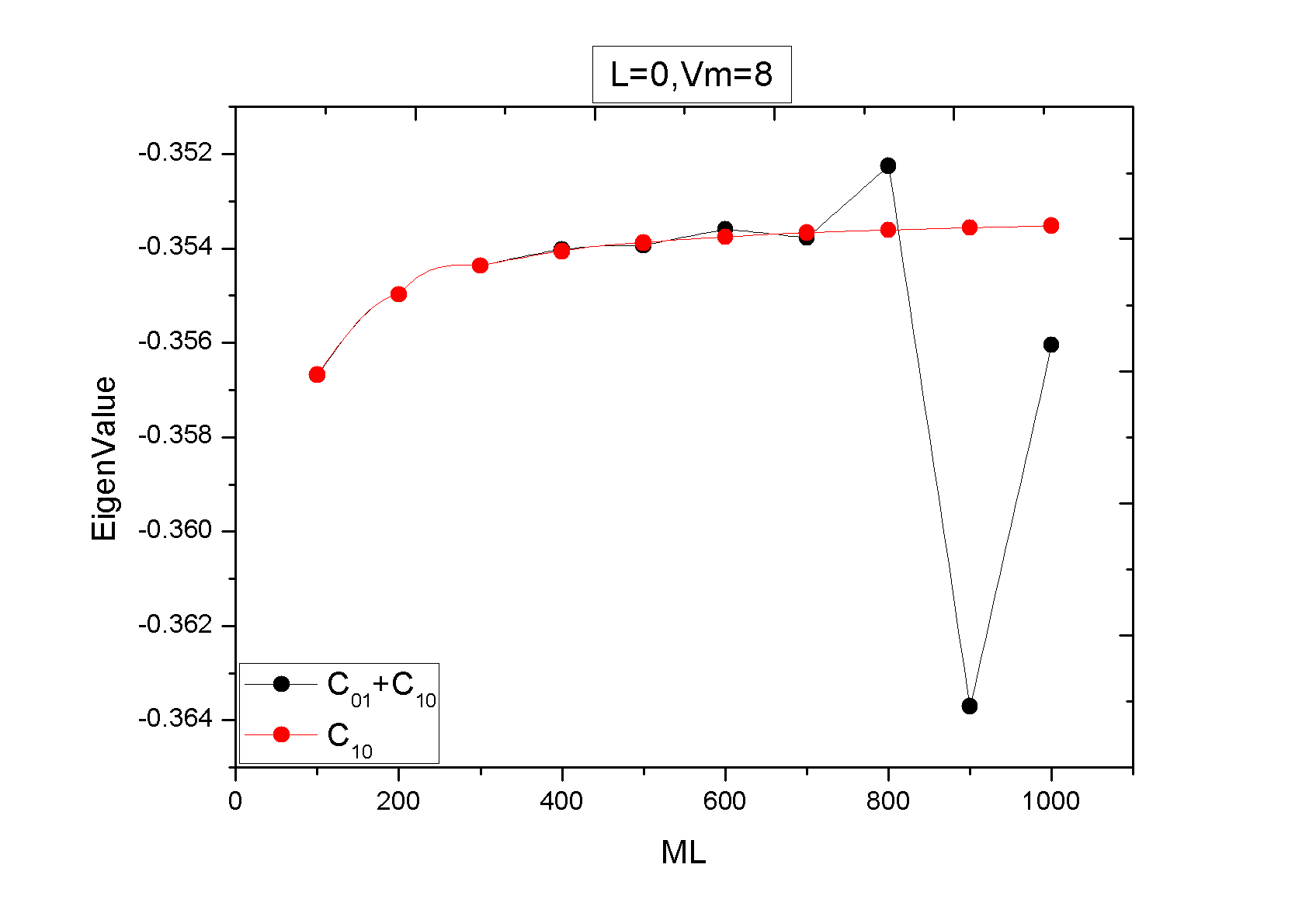}}
\subfigure[Figure 3]{
\includegraphics[width=6.5cm,height=4.5cm]{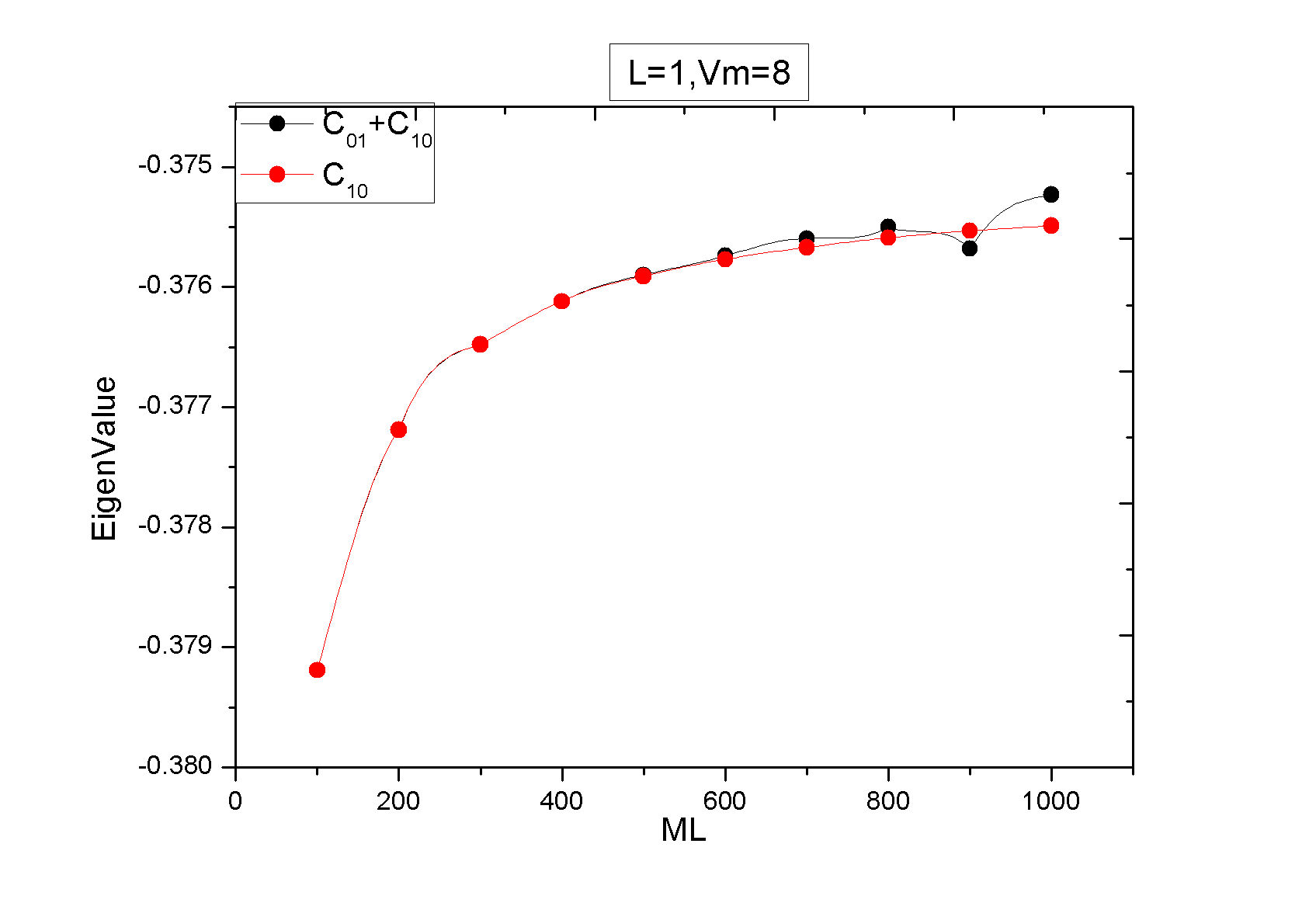}}
\end{figure}\\
Figure 1 plots the case of $l=0,v_m=7v_{i}$, the horizontal axis means the number of grids in each case. The dots in black represent the eigenvalues calculated by the whole operator, while the read ones represent the eigenvalues calculated by merely differential operator. By examining figure 1 to figure 3, the eigenvalues of the whole collision term and the differential term are almost the same. Especially for large $l$ and $v_m$, their differences become much smaller. The oscillations of eigenvalues of the whole collision term may be a result of the influence of integral terms. The result that the eigenvalues are approximately the same applies to other cases as well. Then we go deeper to examine the eigenfunctions corresponding to the minimum eigenvalues calculated respectively by the whole term and only the differential term.\\
\clearpage
\begin{figure}[!h]
\centering
\subfigure[Figure 4]{
\includegraphics[width=6.5cm,height=4cm]{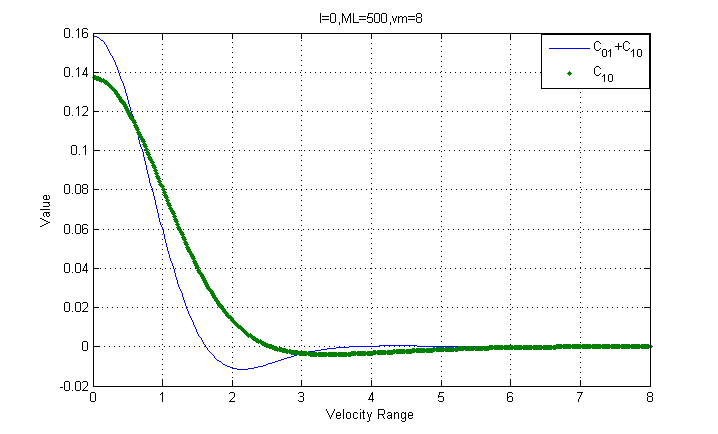}}
\subfigure[Figure 5]{
\includegraphics[width=6.5cm,height=4cm]{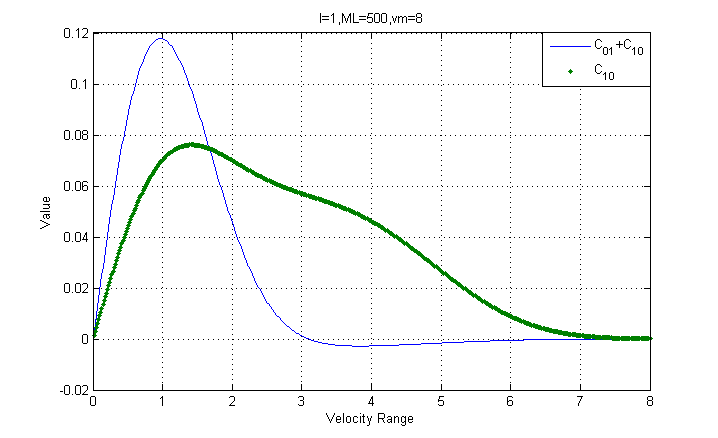}}
\subfigure[Figure 6]{
\includegraphics[width=6.5cm,height=4cm]{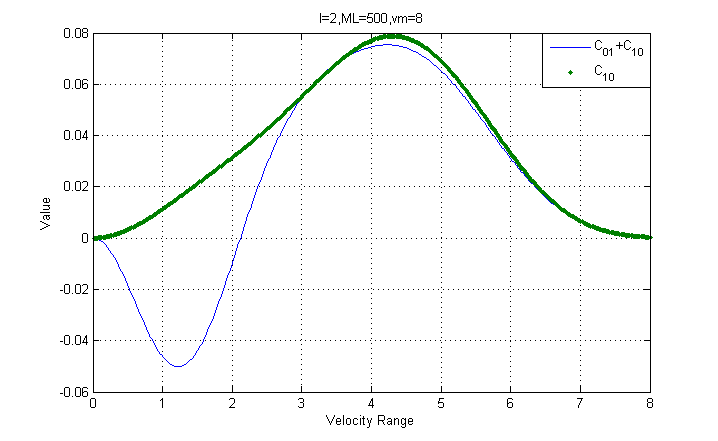}}
\end{figure}
In figure 6,7,8, the maximum velocity is $8v_i$, and the number of girds is 500. The only difference is the subscript $l$. The line with blue color is the corresponding eigenfunction of the minimum eigenvalue of the whole term and the green line is of the differential term. These two kinds of eigenfunctions differ in small velocity but become more alike in the large velocity region. All the eigenfunctions should converge to 0 when $v\rightarrow+\infty$ by considering physical facts. By comparing these 3 graphs, only when $l=0$, the eigenfunctions do not converge to zero when $v$ approaches $0$. For higher order, say $l>0$, all the eigenfunctions start from value of zero. The reason of this phenomenon will be explained in following sections. \\
\section{\Large{Eigenvalues of Differential Operator}}
In this section,we only talk about theoretical properties of differential term in the whole operator. The differential operator have been shown in Eq.\eqref{eq:6}. After neglecting integral operator, the Fokker-Planck equation is like:
\begin{equation}\label{eq:8}
 -i\omega f^1=\hat C_{10}f^1
\end{equation}
To write Eq.\eqref{eq:8} in a more detailed way, we get:
\begin{equation}\label{eq:9}
\begin{split}
 -i\omega f^1&=\Gamma[\frac{n_i}{(2\pi)^{3/2}v_i^3}]^2\big\{-\frac{1}{2v^2}\frac{\partial}{\partial v}[v\frac{\partial H_{10}}{\partial v}e^{-v^2/2}\frac{\partial}{\partial v}(e^{v^2/2}f^1)]\\
 &+\frac{1}{2v^3}\frac{\partial G_{10}}{\partial v}\frac{\partial}{\partial \mu}
 [(1-\mu^2)\frac{\partial f^1}{\partial \mu}]\big\}
\end{split}
\end{equation}
By writing $-i\omega [(2\pi)^{3/2}v_i^3/n_i]^2/\Gamma$ simply as $\lambda$ and then using moment expansion[Eq.\eqref{eq:14}], we gain the eigenvalue equation of the differential operator.
\begin{equation}\label{eq:15}
 \lambda a_l=\frac{l(l+1)}{2v^3}\frac{\partial G_{10}}{\partial v}a_l+\frac{1}{2v^2}\frac{\partial}{\partial v}\big[v\frac{\partial H_{10}}{\partial v}e^{-v^2/2}\frac{\partial}{\partial v}(e^{v^2/2}a_l)\big]
\end{equation}
Eq.\eqref{eq:9} describes the evolution of distribution function according to time perturbed by a test particle. However, the background remains Maxweillian distribution not affected by the test particle. Only considering differential operator costs the equation to lose some basic properties of the whole Rosenbluth collision term such as the conservation of momentum and energy, even though the number of particles still remains conserved. This problem may brings out some unphysical results but in some cases when the differences are not so obvious, the results are still reliable and valid.\\
Eq.\eqref{eq:15} can easily be transformed into Strum-Liouville(SL) type differential equation:
\begin{equation}\label{eq:10}
 \frac{\partial}{\partial v}\big[k(v)\cdot X'(v)\big]-q(v)X(v)+\lambda \rho(v)X(v)=0
\end{equation}
In Eq.\eqref{eq:10}, the specific functions are:
\begin{subequations}
\begin{align}
 X(v)&=-e^{v^2/2}a_l(v)\\
 k(v)&=-ve^{-v^2/2}\frac{\partial H_{10}}{\partial v}\\
 q(v)&=\frac{l(l+1)}{v}e^{-v^2/2}\frac{\partial G_{10}}{\partial v}\\
 \rho(v)&=2v^2e^{-v^2/2}
\end{align}
\end{subequations}
The boundary conditions are:
\begin{equation}
\begin{split}
&\lim_{v\rightarrow 0}X(v)<+\infty\\
&\lim_{v\rightarrow +\infty}X(v)=0
\end{split}
\end{equation}
According Strum-Liouville theorem, when $l=0$, it is easy to prove that all these functions are positive. Thus, all the eigenvalues comprises a complete set and all the eigenfunctions are orthogonal. For orthogonality, the weighing function is $\rho(v)$. Thus, the completeness of the eigenvalues and the orthogonality of the eigenfunctions when $l=0$ are proved rigorously.\\
\section{\Large{Further Discussions on Differential Operator}}
In order to get a deeper insight into differential operator, first we reduce the complex differential equation Eq.\eqref{eq:9} to a simpler form like this:
\begin{equation}\label{eq:20}
 \hat C_{10}a_l=A\cdot\frac{\partial^2 a_l}{\partial v^2}+B\cdot\frac{\partial a_l}{\partial v}+C\cdot a_l=\lambda a_l
\end{equation}
$A,B,C$ are functions of $v$,
\begin{subequations}
\begin{align}
 A&=\frac{1}{2v}\cdot\frac{\partial H_{10}}{\partial v}\label{eq:21}\\
 B&=\frac{1}{2v^2}\big(\frac{\partial H_{10}}{\partial v}+v\frac{\partial^2 H_{10}}{\partial v^2}+v^2\frac{\partial H_{10}}{\partial v}\big)\label{eq:22}\\
 C&=\frac{1}{v}\frac{\partial H_{10}}{\partial v}+\frac{1}{2}\frac{\partial^2 H_{10}}{\partial v^2}+\frac{l(l+1)}{2v^3}\cdot \frac{\partial G_{10}}{\partial v}\label{eq:23}
\end{align}
\end{subequations}
In order to eliminate the first derivative order term in Eq.\eqref{eq:20}, namely $\partial a_l/ \partial v$, a mathematical transformation called "Liouville-Green Transformation" is used:
\begin{equation}\label{eq:24}
 a_l(v)=H(v)\cdot exp(-\frac{1}{2}\int_0^v\xi(v')dv')
\end{equation}
Here we donote $exp(-\frac{1}{2}\int_0^v\xi(v')dv')$ as a new operator $\hat O$, Eq.\eqref{eq:24} can be written in another way:
\begin{equation}
 a_l=H\cdot \hat O=\hat O\cdot H
\end{equation}
After some mathematical procedure, the Eq.\eqref{eq:20} can be expressed in the following form:
\begin{equation}\label{eq:25}
 (\hat O^{-1}\hat C_{10}\hat O)H=A\{H''+[\eta(v)-\frac{1}{4}\xi^2(v)-\frac{1}{2}\xi'(v)]\}=\lambda H
\end{equation}
Where $\xi(v)=B/A,\eta(v)=C/A$. Then, a new operator can be defined: $\hat P=\hat O^{-1}\hat C_{10}\hat O$, its eigenvalue equations is:
\begin{equation}\label{eq:26}
 \hat PH=A\{H''+[\eta(v)-\frac{1}{4}\xi^2(v)-\frac{1}{2}\xi'(v)]\}=\lambda H
\end{equation}
By comparing the form of Eq.\eqref{eq:20} and Eq.\eqref{eq:26}, the eigenvalues are the same for operator $\hat C_{10}$ and the operator $\hat P$. And this is somewhat like an orthogonal transformation by carefully examining the components of $\hat P$.\\
We also define $U(v)$ as a potential like function.
\begin{equation}\label{eq:27}
 U(v)=\eta(v)-\frac{1}{4}\xi^2(v)-\frac{1}{2}\xi'(v)
\end{equation}
This potential like function will determine the property of continuum of the eigenvalues. It will be discussed later. More over, Eq.\eqref{eq:26} are much like a wave equation, and this makes the physical essence more clearer.
\subsection{\large{Behaviors of $U(v)$ when $v\rightarrow0$}}
Because both in $\xi(v)$ and $\eta(v)$, when $v\rightarrow0$ there may be a singularity, it is very necessary to have a thorough discussion about the behaviors of the related functions when $v$ approaches 0.\\
First of all, error function can be expanded using Taylors formula at $v=0$:
\begin{equation}
 erf(x)=\frac{2}{\sqrt{\pi}}(x-\frac{1}{3}x^3+\frac{1}{10}x^5+o(x^7))
\end{equation}
Then function $\partial H_{10}/\partial v$ can be written in progressing form:
\begin{equation}
 \frac{\partial H_{10}}{\partial v}\sim -\frac{8}{3}\pi v-\frac{5}{4}\pi v^3+o(v^5)
\end{equation}
By using Eq.\eqref{eq:0}, we get progressing form of $\partial G_{10}/\partial v$:
\begin{equation}
 \frac{\partial G_{10}}{\partial v}\sim \frac{8}{3}\pi v-\frac{16}{15}\pi v^3+o(v^5)
\end{equation}
By using Eq.\eqref{eq:21}-Eq.\eqref{eq:23}, $A,B,C$ have the form:
\begin{subequations}
\begin{align}
 A&\sim -\frac{4}{3}\pi+o(v^2)\\
 B&\sim -\frac{8}{3}\pi(\frac{1}{v})+o(v)\\
 C&\sim (-4\pi+\frac{8l(l+1)}{15}\pi)+\frac{4l(l+1)}{3}\pi(\frac{1}{v^2})+o(v^2)
\end{align}
\end{subequations}
Finally, we get the progressing behavior of $U(v)$:
\begin{equation}\label{eq:28}
 U(v)\sim (3-\frac{2l(l+1)}{5})-l(l+1)(\frac{1}{v^2})+o(v)
\end{equation}
Eq.\eqref{eq:28} gives some explanations stated about eigenfunctions in section 3.When it is the case that $l=0$, the term with a singularity form $1/v^2$ disappears. Thus, the eigenfunction should not be zero in the $v\rightarrow0$ limit.But when $l>0$, this singularity remains. The potential like function is infinite when $v\rightarrow0$. That is why for $l=1,2$ the eigenfunctions start from 0 to avoid infinity.\\
This phenomenon can also be viewed in a physical way. When $v\rightarrow0$,the ion collision rate becomes infinite, so the perturbation tends to become more isotropic. Hence, the contribution of $P_0$, which describes the isotropic part of distribution function, becomes the largest and the other Legendre Polynomials become less important. So only for $l=0$ the eigenfunction has a totally different behavior in zero velocity limit.
\subsection{\large{Continuum of Eigenvalue Spectrum}}
The continuum of eigenvalues of differential operator can be proved by examining the potential like function $U(v)$. Just like potential barriers and potential walls in quantum mechanics, if the following relation\eqref{eq:29} is satisfied, which means there is no potential barrier to trap the eigenmodes, the eigenvalues will be continuous.
\begin{equation}\label{eq:29}
 U(v)\leq 0
\end{equation}
Here are the graphs for $l=1,2$:
\begin{figure}[!h]
\centering
\subfigure[Figure 7]{
\includegraphics[width=6.5cm,height=4.5cm]{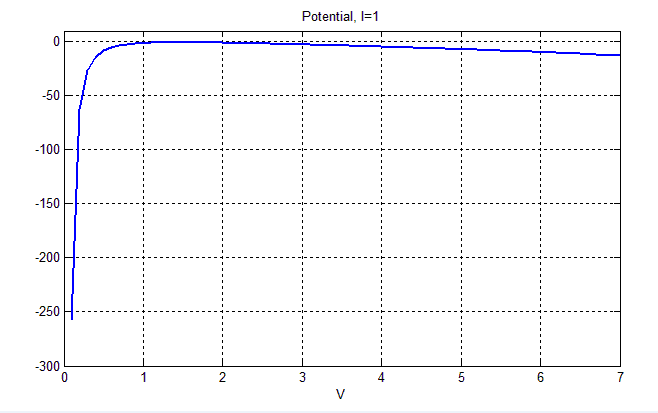}}
\subfigure[Figure 8]{
\includegraphics[width=6.5cm,height=4.5cm]{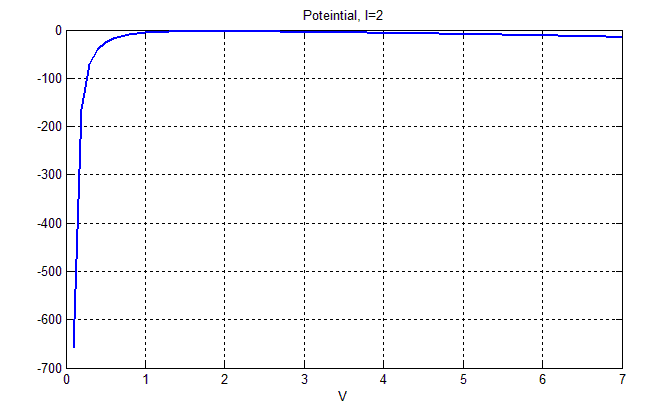}}
\end{figure}
In Figure 7 and Figure 8, the potential like function $U(v)$ is negative which indicates relation\eqref{eq:29} is satisfied. The continuum of the eigenvalues of differential operator also give some evidence that the eigenvalue spectrum of the whole Rosenbluth operator may be continuous as well.
\section{\Large{Conclusion}}
In this paper, firstly we discuss the general collision term of Fokker-Planck equation. Then, by using moment expansion, we get the eigenvalue equation of Rosenbluth collision operator. Some numerical results have been presented, along with the minimum eigenvalues calculated by only using differential operator of the whole operator. These results point out that in some region these two eigenvalue spectrums are almost the same but as to eigenfunctions, there are some differences in low velocity region.\\
A new method on analyzing differential operator is proposed. The eigenvalue equation can be transformed into Strum-Liouville type differential equation. Then the completeness of the eigenvalues and the orthogonality of eigenfunctions when $l=0$ are therefore proved by referring to Strum-Liouville theorem.\\
To further analyze the property of differential operator, Liouville-Green transformation is introduced to reduce the complexity of differential operator. A new operator $\hat P=\hat O^{-1}\hat C_{10}\hat O$ which has the same eigenvalues is defined and all the problems are from the potential like function. Then, we thoroughly examine the behavior of $U(v)$ and get some valid results. The continuum of eigenvalues of the differential term is indicated by the fact that the potential is always below zero.\\
However, some properties of the whole operator are still difficult to interpret because of the integral term which is to complicated to analyze theoretically. More over, some conservation relations are neglected if only differential term is taken into consideration. Hereby, there are still much analytical and numerical work remaining to be done.

\end{document}